\documentstyle[prc,preprint,aps,epsf,array]{revtex}

\def\oln{\overline}

\def\ds{\displaystyle}

\def\pslash#1{\slash \hspace{-2.5mm} #1}
\newcommand{\ignorethis}[1]{}

\def\Omit#1{}

\def\beq{\begin{equation}}
\def\eeq{\end{equation} }
\def\bea{\begin{eqnarray}}
\def\eea{\end{eqnarray}}
\def\eqref#1{Eq.~(\ref{eq:#1})}
\def\eqlab#1{\label{eq:#1}}
\def\figref#1{Fig.~(\ref{fig:#1})}
\def\figlab#1{\label{fig:#1}}

\def\VYP#1#2#3{{\bf #1}, #3 (#2)}  % Volume, page (Year)
\def\NP#1#2#3{Nucl.~Phys.~\VYP{#1}{#2}{#3}}
\def\NPA#1#2#3{Nucl.~Phys.~A~\VYP{#1}{#2}{#3}}

\def\PLB#1#2#3{Phys.~Lett.~B~\VYP{#1}{#2}{#3}}

\def\PRC#1#2#3{Phys.~Rev.~C~\VYP{#1}{#2}{#3}}

\def\PRL#1#2#3{Phys.~Rev.~Lett.~\VYP{#1}{#2}{#3}}

\def\ZP#1#2#3{Z.\ Phys.\  \VYP{#1}{#2}{#3}}

\newcommand{\vslash}[1]{#1 \hspace{-0.5 em} /}
\begin{document}
%\draft
\tighten
\title{Compton scattering in a unitary approach with causality
constraints}
\author{S. Kondratyuk, O. Scholten}
\address{Kernfysisch Versneller Instituut, 9747 AA Groningen, The Netherlands.}
\date{\today}
\maketitle
\begin{abstract}

Pion-loop corrections for Compton scattering are calculated in a novel
approach based on the use of dispersion relations in a formalism
obeying unitarity. The basic framework is presented, including an
application to Compton scattering. In the approach the
effects of the non-pole contribution arising from pion dressing are
expressed in terms of (half-off-shell) form factors and the nucleon
self-energy. These quantities are constructed through the application
of dispersion integrals to the pole contribution of loop diagrams, the
same as those included in the calculation of the amplitudes through a
K-matrix formalism. The prescription of minimal substitution is used
to restore gauge invariance. The resulting relativistic-covariant
model combines constraints from unitarity, causality, and crossing
symmetry.

\end{abstract}

\vspace{2cm}
\noindent
{\bf Key Words}:
Nucleon-photon vertex, Off-shell form factors, K-matrix formalism, Compton scattering,
Dispersion relations.

\noindent {\bf 1999 PACS}:
11.55.Fv, 13.40.Gp, 13.60.Fz

%\bigskip
\Omit{
\noindent
Corresponding author: \\
O. Scholten, Kernfysisch Versneller Instituut \\
Zernikelaan 25, 9747 AA Groningen, The Netherlands  \\
e-mail:  Scholten@KVI.NL, \\
phone:  +31-(0)50-363-3552, fax:    +31-(0)50-363-4003 \\
}
- - - - - - - - - - - - - - - - -  \today\ - - - - - - - - - - - - - - - - -

\pacs{11.55.Fv, 13.40.Gp, 13.60.Fz}

\section{Introduction}

In Ref.~\cite{Kon99} a relativistic covariant model was presented for
pion-nucleon scattering in which constraints due to unitarity were
taken into account through the use of the K-matrix formalism with a
non-perturbative dressing of the $\pi N N$-vertex. An approach based
on the use of dispersion relations was employed in this dressing. It
was shown that as a result of the dressing the effective form factors
were softened. Here we extend this work to include processes involving
photons. The fact that our procedure is based on the use of dispersion
relations, thus incorporating analyticity, gives a considerable
advantage. It is known\cite{Ber93,Lvo97,Hun97,Han98} that in photon induced
processes important constraints are imposed by the condition that the
amplitude for the process has to be analytic, especially at energies
near the pion-production threshold. The usage of dispersion relations
allows for an implementation of these analyticity constraints in a
unitary K-matrix approach. In quantum field theory analyticity is
based on the condition of causality and is a fundamental property of
the S-matrix.

As with any process involving photons one should take care to obey
gauge invariance of the amplitude since otherwise low-energy theorems
may be violated. In the present approach we used the minimal
substitution procedure. We discuss this method in some detail and
present general formulas.

In the procedure the effects of dressing are expressed in terms of
form factors and self-energies. The present work is focused on the
$\gamma N N$-vertex. Electromagnetic vertices of the nucleon with one
or both nucleons off-shell have been studied in the past. The method
of dispersion relations was applied in Refs.\cite{Bin60}. Dynamical
models based on a perturbative dressing of the vertex with meson
loops, within effective Lagrangian approaches, were developed in
Refs.\cite{Nau87}. The role of off-shell nucleon-photon form factors
has been investigated, for example, in models for proton-proton
bremsstrahlung \cite{Nym71} and virtual Compton
scattering \cite{korschjong}. Analyticity considerations have been
used in  Refs.\cite{Ber93,Lvo97,Hun97,Han98} to construct
amplitudes for Compton scattering from those of pion photoproduction
through the application of dispersion relations. The present
approach incorporates dispersion relations (or analyticity
considerations) in a more microscopic approach to pion and photon
induced reactions on the proton.

\Omit{In the present Non-perturbative, Gauge-invariant, Relativistic-co-variant model obeying
Analyticity and Unitarity constraints (NGRAU-model)}

The present model consists of two stages implemented in an iteration
procedure to reach self-consistency. In the first stage effective
two-, three- and four-point Green's functions are built which
incorporate non-perturbative dressing due to non-pole parts of loop
diagrams. At each iteration step, the imaginary parts of the loop
integrals are found by applying Cutkosky rules \cite{cut}. In doing
so, only the intermediate states with one nucleon and one pion are
taken into account. The real parts are constructed using dispersion
relations \cite{Bin60}. The dispersion integrals converge due to the
sufficiently fast falloff of the $\pi N N$ form factors in the loop
diagrams. The resulting $\gamma N N$-vertex is normalized in such a
way that, at the point where both nucleons are on-shell, it reproduces
the physical anomalous magnetic moment of the nucleon.

In the second stage a K-matrix formalism \cite{gouds,korsch,Kor98} is
employed to calculate the T-matrix, where the kernel, the K-matrix, is
built from tree-level diagrams using the dressed vertices and
propagators calculated in the first stage. Through the use of the
K-matrix formalism the pole contributions of loop integrals are taken
into account. The T-matrix obtained from thus constructed K-matrix
will contain the principal-value parts of the same loop integrals
which were included in the dispersion calculation for the form factors
and self-energies, implementing analyticity in the K-matrix framework.
Since the dressing is formulated in terms of effective vertices and
propagators through the use of form factors and self-energies, a
broader application might be possible.

The $\gamma N N$-vertex must satisfy the Ward-Takahashi identity.
This is achieved in our model by including a loop
diagram with a four-point $\gamma \pi N N$-vertex (the contact term).
The latter is constructed based on the dressed $\pi N N$-vertex using
the minimal substitution prescription (various constructions of
contact terms can be found in Refs.\cite{Oht89,Gro87}). Such a
procedure leads to a unique result only for the longitudinal (with
respect to the photon momentum) part of the four-point
vertex. To investigate the role of the transverse terms, we calculated
the electromagnetic form factors utilizing two different $\gamma \pi N
N$ vertices. To provide current conservation in the description of
Compton scattering also a contact $\gamma \gamma N N$ term is built
using the minimal substitution prescription.

The model is geared to the calculation of pion-photoproduction and
Compton scattering on the nucleon. To study effects of the dressing we
compare the $f_{EE}^{1-}$ partial wave amplitude for Compton scattering obtained using the
dressed $\gamma N N$ and $\pi N N$ vertices and nucleon propagator
with that of a calculation using bare vertices and the free
propagator. We also calculated the electric polarizability of the proton.

In Section II we outline the construction of the K-matrix in a
coupled-channel unitary description of Compton scattering, pion
photoproduction and pion-nucleon scattering. Details of the dressing
procedure are given in Sections II.A and II.B where special attention
is payed to vertices with photons. Numerical results
on $\gamma NN$-vertices, expressed in terms of half-off-shell form
factors, are given in Section III. In Section IV the formalism is
applied to Compton scattering where we focus on the effects of
dressing on observables. Conclusions are given in Section V.

\section{Structure of the K-matrix}

Our model is based on the K-matrix formalism\cite{gouds,korsch,Kor98}
and to explain our procedure we work in a simple model space with only
the nucleon, pion and photon degrees of freedom. Only the one-pion
threshold discontinuities are taken into account explicitly.

To describe simultaneously pion-nucleon scattering,
pion photoproduction and Compton scattering the scattering matrix
has two indices corresponding to the channel in the initial and
final state, ${\mathcal{T}}_{c' c}$, where the indices can be
$\pi$ or $\gamma$ for the channels $\pi N$ or $\gamma N$, respectively.
The Bethe-Salpeter equation for the scattering matrix can be written as
\beq
{\mathcal{T}}_{c' c}\,=\,V_{c' c}\, +\sum_{c''}V_{c' c''}\,
{\mathcal G}_{c''}\,{\mathcal T}_{c'' c}\;,
\eqlab{k1}
\eeq
where $V_{c' c}$ is the sum of irreducible diagrams describing the process
$c \rightarrow c'$ and ${\mathcal{G}}_{c''}$ is the two-body
propagator pertinent to the channel $c''$. ${\mathcal{G}}_{c''}$ contains
the pole contribution $i\delta_{c''}$ which is imaginary, according to
Cutkosky rules, and the regular (principal-value) part
${\mathcal {G}}^R_{c''}$ which is real,
\beq
{\mathcal{G}}_{c''}\,=\,{\mathcal{G}}^R_{c''}\,+\,i\delta_{c''}\;.
\eqlab{k2}
\eeq
The K-matrix can be introduced as the solution of the equation
\beq
K_{c' c}\,=\,V_{c' c}\,+\,\sum_{c''}V_{c' c''}\,{\mathcal{G}}^R_{c''}\,
K_{c'' c}\;.
\eqlab{k3}
\eeq
According to this formula, the loop diagrams contributing to the K-matrix
contain only the principal-value part of the two-particle propagator.
We assume throughout that $V$ is a sum of tree-level diagrams. The
remaining pole contribution enters explicitly in the equation for the
T-matrix expressed in terms of the K-matrix,
\beq
{\mathcal{T}}_{c' c}\,=\,K_{c' c}\,+\,\sum_{c''}K_{c' c''}\,i\delta_{c''}\,
{\mathcal{T}}_{c'' c}\; ,
\eqlab{k4}
\eeq
which can be obtained from Eqs.(\ref{eq:k1}-\ref{eq:k3}).
A formal solution of \eqref{k4} can be written as (suppressing the channel indices)
\beq
{\mathcal T}= K\,(1-K i\delta)^{-1} \;,
\eqlab{k5}
\eeq
from which it follows that the S-matrix,
$S=1+2i\mathcal{T}$, will be unitary provided $K$ is hermitian.

\eqref{k3} suggests an interpretation of the K-matrix in terms of a
dressing of a potential $V_{c' c}$ with principal-value parts of loop
integrals. To illustrate this for the case of Compton scattering we
choose  $V_{\gamma \gamma}$ and $V_{\gamma \pi}$
as the sum of  s- and u- and t-channel tree diagrams
plus a possible four-point vertex,
where the free nucleon propagator and bare nucleon-photon vertices are
used. Up to second order in $V_{c' c}$ and leading order in the
electromagnetic coupling constant, $K_{\gamma \gamma}$ can be written as
\beq
K^{(2)}_{\gamma \gamma}=V_{\gamma \gamma}+V_{\gamma \pi}\,
{\mathcal{G}}^R_{\pi}\,V_{\pi \gamma}\;.
\eqlab{k6}
\eeq
The set of diagrams corresponding to the right-hand side of this
equation is depicted in \figref{f11}. The notation $ss$, $su$ etc. for
the loop diagrams refer to their structure in terms of the s-, u- ,
t-channel and contact tree diagrams contributing to $V_{\gamma \pi}$.
The index ${\it Re}$ at the loops indicates that only the principal-value
integrals are taken into account, in accordance with \eqref{k6}.
Consequently, the self-energy functions and form factors parametrizing
these loops are real functions. One can see that the one-particle
reducible diagrams in \figref{f11} (diagrams {\it ss, su, st, sc, us,
uc, ts, cs} and {\it cu}) are part of a dressing of the nucleon
propagator and half-off-shell nucleon-photon vertices. The other
diagrams in \figref{f11}, which are one-particle irreducible, are
necessary to ensure the gauge invariance of $K^{(2)}_{\gamma \gamma}$.

The above description of $K^{(2)}_{\gamma \gamma}$ serves as an
introduction to the dressing procedure described below. In the full
model dressing up to infinite order is taken into account, expressed
in terms of an integral equation. Gauge invariance is maintained
through the introduction of an appropriate contact term. It should be
pointed out that \eqref{k3} dictates that only principal-value parts
of the loop integrals (or, equivalently, only the real parts of the
form factors and self-energy functions) are taken into account in the
iteration procedure for the vertices and the nucleon propagator.

To summarize, in our procedure we construct the $K$ matrix which
enters in \eqref{k5} as the sum of tree-level diagrams (those for
$K_{\gamma \gamma}$ and $K_{\gamma \pi}$ are depicted in
\figref{dia-K1}) where dressed nucleon propagators, dressed
nucleon-pion, dressed nucleon-photon vertices, and contact terms (for
gauge invariance) are used. The use of dressed quantities is implied
by \eqref{k3}, where, due to taking the principal value integrals, the
effect of dressing can be expressed in terms of purely {\em real} form
factors or self-energy functions. These real functions are obtained by
applying dispersion relations to the one-particle reducible pole
contributions from \eqref{k4}, thereby implementing analyticity
(causality) constraints in the calculation of the T-matrix. The
procedure followed is discussed in detail in the following sections.

\subsection{The dressing procedure \label{sec:III}}

The most general $\gamma NN$-vertex for a real photon with momentum
$q=p'-p$,
in which the outgoing nucleon is on the mass shell, $p^{\prime
2}=m^2$, can be written\footnote{The notation of
Ref.\cite{Bjo64} is used throughout this paper.} as\cite{Bin60}
\beq
e \Gamma_{\mu}(p) = e \sum_{l={\pm}} \left\{ \gamma_{\mu} F_1^{l}(p^2) +
i \frac{\sigma_{\mu \nu}q^{\nu}}{2m} F_2^{l}(p^2)
\right\}\,\Lambda_l(p)\;,
\eqlab{finon}
\eeq
where $e$ and $m$ are the elementary electric charge
and the mass of the nucleon and
\beq
\Lambda_{\pm}(p) \equiv \frac{{\pm}\vslash{p}+m}{2m} \;.
\eqlab{proj}
\eeq
The isospin structure of the form factors is taken as $F=F^s + \tau_3
F^v$. The dressing of this vertex is expressed in terms of a system
of integral equations, shown diagrammatically in \figref{f1},
\beq
\Gamma_{\mu,R}(p) = \Gamma_{\mu}^0(p) +
\mbox{D.I.} \Big\{ \, \Gamma_{\mu,I}[1]+
\Gamma_{\mu,I}[2] + \Gamma_{\mu,I}[3]\, \Big\} \;,
\eqlab{sys}
\eeq
where ``D.I.'' implies taking a dispersion integral,
$\Gamma_{\mu,R}$ contains only the real parts of the form factors and
each of the terms will be discussed in detail in the following.
This equation expresses the dressing of a
bare vertex $\Gamma_{\mu}^0(p)$ with an infinite series of pion
loops. The bare vertex is taken as
\beq
\Gamma_{\mu}^0(p) =
( \gamma_{\mu} \hat{e}_N + i \hat{\kappa}_B
\frac{{\sigma}_{\mu \nu} q^{\nu}}{2 m} ),
\eqlab{ver_bare}
\eeq
where $\hat{e}_N = (1+{\tau}_3)/2$ and $\hat{\kappa}_B = \kappa_B^s
+ \tau_3 \kappa_B^v$ is the bare anomalous magnetic moment of
nucleon, adjusted to provide the normalization
\eqref{normalF2} of the dressed vertex.
The solution of \eqref{sys} is obtained by requiring self-consistency
in an iteration procedure. We consider irreducible vertices, which implies that the
external propagators are not included in the dressing of the vertices.

Every iteration step proceeds as follows. The imaginary or pole
contributions of the loop integrals for both the propagators and the
vertices are obtained by applying cutting rules \cite{cut}. Since the
outgoing nucleon is on-shell, the only kinematically allowed cuts are
those shown in \figref{f1}. In calculating these pole contributions,
we retain only real parts of the form factors and nucleon
self-energies from the previous iteration step as required by
\eqref{k3}. We note that all integrals on the right-hand of
\eqref{sys}, except the one over $\Gamma_{\mu,I}[2]$, are
inhomogeneities of the equation because they do not depend on the
$\gamma N N$-vertex. Therefore, they need to be calculated only once.
The dressed $\pi NN$-vertices and the nucleon propagator are taken
from Ref.\cite{Kon99} where they have been constructed using a
compatible procedure.

The real parts of the form factors are calculated at every iteration
step by applying dispersion relations \cite{Bin60} to the imaginary
parts just calculated.

This procedure is repeated to obtain a converged solution. The
convergence criterion is imposed for a normalized root-mean-square
difference $d_n$ for the form factors between two subsequent iteration
steps $n$ and $n+1$. The convergence criterion is that $d_n < 10^{-8}$
for a large number of iterations.

One of the advantages of the use of cutting rules is that throughout
the solution procedure we need vertices with only one virtual nucleon
(half-off-shell vertices), as can be seen from \figref{f1}. In other
words, the knowledge of {\it full}-off-shell form factors will not be
required for the calculation of the pole contributions to the loop
integrals. Also for the construction of the K-matrix only
half-off-shell vertices are required.

Since in the dressing of the $\pi NN$-vertex a bare form factor is
required for regularization, the described procedure obeys analyticity
only approximately (the bare $\gamma NN$-vertex does not contain form
factors see \eqref{ver_bare}). The influence of the singularities of
the bare form factor can be diminished in the kinematical region of
interest by a rather large width. This is
consistent with the fact that the bare form factor represents physics left
out from the model and thus should vary at an energy scale larger than
the heaviest meson included explicitly.

\subsection{The loop integrals}

The pole
contribution $\Gamma_{\mu, I}[1]$, of the first loop integral on
the right hand side of \eqref{sys} comes
from cutting the nucleon propagator
$S(p-k)$ and the pion propagator $D(k^2)$, i.e.\ from putting the
corresponding particles on their mass-shell (see \figref{f1}).
According to Cutkosky rules \cite{cut}, we replace $S(p-k)$ with
$-2 i \pi (\vslash{p}-\vslash{k}+m) \delta((p-k)^2-m^2) \Theta(p_0-k_0)$
and $D(k^2)$ with $-2 i \pi \delta(k^2-m_{\pi}^2) \Theta(k_0)$, where
$m_\pi$ is the pion mass.

The half-off-shell pion-nucleon vertex for an incoming nucleon with momentum  $p$ and
an on-shell outgoing nucleon ($p^{\prime}$) entering in the
expressions is written as
\begin{equation}
\Gamma_{5,\alpha}(p) =
\tau_\alpha \Gamma_5(p) =
\tau_{\alpha}\,\gamma_5 \Big[ G_1(p^2)+
\frac{\vslash{p}-m}{m} G_2(p^2)\Big] \;,
\eqlab{pinnver}
\end{equation}
where $k=p-p^{\prime}$ is the momentum of the pion. The functions
$G_{1,2}(p^2)$ are
the (half-off-shell) form factors in the nucleon-pion vertex.
In the approach adopted in \cite{Kon99} we find that the dependence of
the form factor on the pion momentum is small which is therefore ignored.
The pion-nucleon coupling constant is taken from Ref.~\cite{Kor98}, $g=13.02$.

Denoting $g_i \equiv Re G_i$, the pole contribution reads
\begin{eqnarray}
\Gamma_{\mu, I}[1] &=& {\displaystyle \frac{2 \tau_3}{8 \pi^2} \int }\! \,
d^4 k\,\gamma_5\, g_1(m^2)\, (\vslash{p}-\vslash{k}+m)\, \gamma_5
\Big[\, g_1(p^2)+
{\displaystyle \frac{\vslash{p}-m}{m}} g_2(p^2)\, \Big]\,
{\displaystyle \frac{2 k_{\mu} + q_{\mu}}{(k+q)^2-m_{\pi}^2}} \nonumber \\
&&\times \delta((p-k)^2-m^2)\,
\Theta(p_0-k_0)\, \delta(k^2-m_{\pi}^2)\, \Theta(k_0) \;, \eqlab{pole1}
\end{eqnarray}
where $m_\pi$ is the pion mass.
The pion-photon vertex $\Gamma_{\mu,\alpha \beta}(k^{\prime},k)$ is chosen such that
it satisfies the Ward-Takahashi identity with the free pion propagator,
\beq
\Gamma_{\mu,\alpha \beta}(k^{\prime},k)=(\hat{e}_{\pi})_{\alpha \beta}
(k_{\mu}+k^{\prime}_{\mu}) \;,
\eeq
where the pion charge operator
$(\hat{e}_{\pi})_{\alpha \beta} = -i \epsilon_{\alpha \beta 3}$.
\Omit{The isospin factor $2\tau_3$ appears in \eqref{pole1} due to the isospin structure of
the the pion-nucleon vertices.}

Using the notation introduced in \eqref{basis}, one can write
\beq
\Gamma_{\mu, I}[1]={\displaystyle \sum_{i=1}^{4}} c^i[1]\, (e_i)_{\mu}
\eeq
where it has implicitly been assumed that the final nucleon is on the
mass-shell and where
\bea
c^i[1]&=&{\displaystyle \frac{r(p^2)}{16 \pi p^2}\,
\Theta(p^2-(m+m_{\pi})^2)
\int_{-1}^1\! d x\,{V}^i(x)\,
\frac{g_1(m^2)}{(k+q)^2-m_{\pi}^2} },
\eqlab{FF1x} \\
V^i(x)&=&\tau_3 {\displaystyle \sum_{j=1}^{6}}(E^{-1})^i_j\,\Bigg{\langle}\,
(\theta^j)^{\mu}\mbox{\Large ,}\; \Lambda_{+}(p^{\prime})\,
 (-\vslash{p}+\vslash{k}+m)\,
\big[\, g_1(p^2) \nonumber \\
&&+ {\displaystyle \frac{\vslash{p}-m}{m}} g_2(p^2)\, \big]
(2 k_{\mu} + q_{\mu})\, \Bigg{\rangle} \eqlab{vbig}
\eea
where the brackets ${\langle} \cdot , \cdot {\rangle}$ are
defined in \eqref{dual}, the  $(\theta^j)^{\mu}$ is the basis in
the dual space,
$q=p'-p$ and
$r(p^2)=\sqrt{\lambda(p^2,m^2,m_{\pi}^2)}$, with the K\"all\'en function
$\lambda(x,y,z)\equiv (x-y-z)^2-4yz$.
The integral in \eqref{FF1x} is a Lorentz-scalar and therefore can be evaluated in any frame of
reference. We choose the rest frame of the incoming nucleon, i.e.\ we put
$p_{\mu} = W\,\delta_{\mu 0}$, where $W=\sqrt{p^2}$ is the invariant mass of
the off-shell nucleon. Furthermore we introduced $x$, the cosine of the
polar angle between the three-vectors $(-\overrightarrow{q})$ and
$\overrightarrow{k}$. The integral in \eqref{FF1x} is done numerically.

The term  $\Gamma_{\mu,I}[2]$
depends on the unknown half-off-shell $\gamma N N$-vertex and therefore has to
be considered in the context of the iteration procedure. As explained
above, when calculating $\Gamma_{\mu,I}^{n+1}[2]$, the pole
contribution to the n+1$^{st}$ iteration
for $\Gamma_{\mu}[2]$, we retain only
the real parts of $F^{\pm,n}_i(p^2)$ from the previous iteration as
well as of the nucleon-pion form factors
and the functions $\alpha(p^2)$ and $\xi(p^2)$ parametrizing the
renormalized dressed nucleon propagator
\beq
S(p)  = \Big[ \alpha(p^2) \Big( \vslash{p}-\xi(p^2) \Big) \Big]^{-1}
\;.
\eqlab{prop-inv}
\eeq
The functions $\alpha(p^2)$, $\xi(p^2)$, as well as
$G_{1,2}(p^2)$, were calculated in  Ref.\cite{Kon99}.
Using the same approach as for $\Gamma_{\mu,I}[1]$ we write
\beq
\Gamma_{\mu,I}^{n+1}[2]={\displaystyle \sum_{i=1}^{4}}
c^{i,n+1}[2]\, (e_i)_{\mu}
\eeq
where
\bea
c^{i,n+1}[2]&=&\frac{r(p^2)}{32 \pi p^2}\,\Theta(p^2-(m+m_{\pi})^2)
\int_{-1}^1\!dx\,
\frac{{U}^i(x)}{\alpha((p^\prime-k)^2)[(p^\prime-k)^2-
\xi^2((p^\prime-k)^2)]},
\eqlab{FF2x}\\
U^i(x)&=&\sum_{j=1}^{6} (E^{-1})^i_j \Bigg\langle\;(\theta^j)^\mu
\mbox{\Large ,}\,\Lambda_{+}(p^\prime)\Bigg\{ \tau_\alpha
\gamma_5 \Big[ g_1((p^\prime-k)^2)+
{\ds \frac{\vslash{p}^\prime-\vslash{k}-m}{m}}g_2((p^\prime-k)^2) \Big]
\nonumber \\
&&\times \Big(\vslash{p}^\prime-\vslash{k}+\xi((p^\prime-k)^2)\Big) \Bigg\}\,
\Bigg\{ \Lambda_{+}(p^\prime-k) \Big[ \gamma_\mu
f^{+}_1((p^\prime-k)^2) \nonumber \\
&&+ i{\ds \frac{\sigma_{\mu \nu}q^\nu}{2m}}
f^{+}_2((p^\prime-k)^2) \Big]
+ \Lambda_{-}(p^\prime-k)\Big[ \gamma_\mu
f^{-}_1((p^\prime-k)^2) + i{\ds \frac{\sigma_{\mu \nu}q^\nu}{2m}}
f^{-}_2((p^\prime-k)^2) \Big]\Bigg\}  \nonumber \\
&&\times (\vslash{p}-\vslash{k}+m)\tau_\alpha \gamma_5
\Big[ g_1(p^2)+{\ds \frac{\vslash{p}-m}{m}}g_2(p^2) \Big] \Bigg\}\;
\Bigg\rangle \;,  \eqlab{ubig}
\eea
where $f^{\pm}_i(p^2) \equiv Re (F^{\pm})^n_i(p^2)$. At the required
kinematics the functions $\alpha(p^2)$ and $\xi(p^2)$ are real.
Note that the $f$'s contain isospin operators.

The term $\Gamma_{\mu,I}[3]$ contains a ``contact"
$\gamma \pi N N$-vertex. We build such a vertex
applying the procedure of minimal substitution (see Appendices B and C for
details) to the dressed half-off-shell $\pi N N$-vertex,
\begin{eqnarray}
&&\Big(\Gamma_{\gamma \pi NN}\Big)_\alpha^\mu(p^\prime,p,q) = \nonumber \\
&&-\tau_\alpha\hat{e}\Bigg\{ {\ds \frac{2p^\mu+q^\mu}
{(p+q)^2-p^2}}\big[ \Gamma^5(p+q)-\Gamma^5(p) \big] + \gamma^5
{\ds \frac{g_2((p+q)^2)}{m}}\big[ \gamma^\mu - \vslash{q}\,
{\ds \frac{2p^\mu+q^\mu}
{(p+q)^2-p^2}} \big] \Bigg\}  \nonumber \\
&& - \hat{e}\tau_\alpha\Bigg\{ {\ds \frac{2p^{\prime \mu}-q^\mu}
{(p^\prime-q)^2-p^{\prime 2}}}\big[ - \overline{\Gamma^5}(p^\prime-q)+
\overline{\Gamma^5}(p^\prime) \big] +
\big[ \gamma^\mu + {\ds \frac{2p^{\prime \mu}-q^\mu}
{(p^\prime-q)^2-p^{\prime 2}}}\,\vslash{q} \big]{\ds \frac{g_2((p^\prime-q)^2)}
{m}}\,\gamma^5 \Bigg\}, \eqlab{gampinn1}
\end{eqnarray}
where Eqs.~(\ref{eq:contPS},\ref{eq:contPV1}), $p'=p+q$ and \eqref{pinnver} with
$\overline{\Gamma^5}(p)=\gamma_0 \Big( \Gamma^5(p) \Big)^\dagger
\gamma_0$ have been used.
Using the same notation as before we obtain
\beq
\Gamma_{\mu,I}[3]={\displaystyle \sum_{i=1}^{4}}
c^{i}[3]\, (e_i)_{\mu} \;,
\eeq
with
\bea
 c^i[3]&=&-\frac{r(p^2)}{64\pi p^2}\,\Theta(p^2-(m+m_{\pi})^2)\,
\Bigg[(3-\tau_3) \int_{-1}^1 \!dx{W}^i_1(x)
+3(1+\tau_3)\int_{-1}^1\!dx{W}^i_2(x)\Bigg] \;,
\eqlab{FF3} \\
W^i_1(x)&=&\sum_{j=1}^{6} (E^{-1})^i_j \Bigg\langle\,
(\theta^j)^\mu \mbox{\Large ,}\,\Lambda_{+}(p^\prime)
 \Bigg\{(2p_\mu-2k_\mu+q_\mu)\Big[
{\ds \frac{g_{12}((p'-k)^2)-g_{12}(m^2)}{(p'-k)^2-m^2)}} \nonumber \\
&&-{\ds\frac{\vslash{p}-\vslash{k}}{m}\,\frac{g_2((p'-k)^2)-g_2(m^2)}
{(p'-k)^2-m^2}\Big]-\frac{\gamma_\mu}{m}}g_2((p'-k)^2)\Bigg\}
(m-\vslash{p}+\vslash{k})  \nonumber \\
&&\times \Big\{g_{12}(p^2)+{\ds \frac{\vslash{p}}{m}}g_2(p^2)\Big\}
\Bigg\rangle \;, \eqlab{wbig1}\\
W^i_2(x)&=&\sum_{j=1}^{6} (E^{-1})^i_j \Bigg\langle\,
(\theta^j)^\mu \mbox{\Large ,}\,\Lambda_{+}(p^\prime)
\Bigg\{ (2p^\prime_\mu-q_\mu) \Big[{\ds \frac{g_{12}(p^2)-
g_{12}(m^2)}{p^2-m^2}} \nonumber \\
&&+{\ds \frac{g_2(p^2)-g_2(m^2)}{p^2-m^2}\,
\frac{\vslash{p}^\prime}{m}}\Big]
+{\ds \frac{\gamma_\mu}{m}}g_2(p^2)\Bigg\}
(m-\vslash{p}+\vslash{k})  \Big\{g_{12}(p^2)+
{\ds \frac{\vslash{p}}{m}}g_2(p^2)\Big\} \Bigg\rangle \;, \eqlab{wbig2}
\eea
where $g_{12}\equiv g_1-g_2$, and $r(p^2)$ is defined as in \eqref{vbig}.
An alternative expression
for $\Gamma_{\mu, I}[3]$ is obtained if, instead of \eqref{contPV1},
one uses the contact term \eqref{contPV2}. The choice of the contact term
has an influence on the nucleon-photon form factors,
as will be described below.

Since the half-off-shell form factors are analytic \cite{Bin60} in the
complex $p^2$-plane with the cut from
the pion threshold $(m+m_{\pi})^2$ to infinity, dispersion
relations can be used to construct the real parts from the imaginary parts.
In our model the imaginary parts of the form factors $F_2^\pm$ vanish at infinity.
At every iteration step we thus write
\beq
\mbox{Re} F_2(p^2)=\hat{\kappa}_B+\frac{\mathcal{P}}{\pi}
\int_{(m+m_{\pi})^2}^{\infty}\!dp^{\prime 2}\,
\frac{\mbox{Im} F_2(p^{\prime 2})} {p^{\prime 2}-p^2},
\eqlab{drF2}
\eeq
where we have dropped the superscripts $^{\pm}$  of the form factors
to keep the expression transparent.  The constant $\hat{\kappa}_B$
originates from
the first term on the right-hand side of \eqref{sys}. Note that,
according to \eqref{ver_bare}, $\kappa_B^{s,v}$ are chosen the same for the
form factors $F_2^{+}(p^2)$ and $F_2^{-}(p^2)$. They are fixed
by the requirement that the vertex reproduces the physical anomalous
isoscalar and isovector magnetic
moment when both nucleons are on-shell,
\beq
F_2^{+,s}(m^2)=-0.06 \mbox{\ \ \ and\ \ }
F_2^{+,v}(m^2)=1.85 \;.
\eqlab{normalF2}
\eeq

In terms of the
parametrization of the propagator, \eqref{prop-inv}, the
Ward-Takahashi identity \eqref{wti_gen} gives
\beq
(F_1^{-})^{s,v}(p^2)=\frac{\alpha(p^2)}{2}
\eqlab{wtiFm}
\eeq
and
\beq
(F_1^{+})^{s,v}(p^2)=\frac{\alpha(p^2)m}{p^2-m^2}\Big[
\frac{p^2+m^2}{2m}-\xi(p^2) \Big] \;.
\eqlab{wtiFp}
\eeq
where $\lim_{p^2\rightarrow m^2}(F_1^{+})^{s,v}(p^2)$ is finite
because $\lim_{p^2\rightarrow m^2}\xi(p^2)=m$ due to the correct location of
the pole of the renormalized propagator. In Ref.\cite{Kon99} the loop
contribution to the self-energy vanishes  in the limit $p^2 \rightarrow
\infty$ and therefore
$\lim_{p^2\rightarrow \infty}(F_1^\pm)^{s,v}(p^2)-Z_2/2=0$.
$Z_2$ is the nucleon-field renormalization constant.
The form factors $(F_1^{\pm})^{s,v}(p^2)$ thus obey the dispersion
relations (omitting the superscripts)
\beq
\mbox{Re} F_1(p^2)=\frac{Z_2}{2}+\frac{\mathcal{P}}{\pi}
\int_{(m+m_{\pi})^2}^{\infty}\!dp^{\prime 2}\,
\frac{\mbox{Im} F_1(p^{\prime 2})}{p^{\prime 2}-p^2} \;,
\eqlab{drF1}
\eeq
applied at every iteration step.
%
%The constant $Z_2$ found in \cite{Kon99} equals $0.85$.

\section{The form factors}

The results presented in this section are obtained using the $\pi N
N$-vertex \eqref{pinnver} and the nucleon propagator calculated in
Ref.\cite{Kon99}. The solution for the nucleon-pion form factors
depends on the choice of the bare vertex. However, the half-width of
this form factor should not exceed a rather well defined maximum for
the procedure to converge. For the present calculation of the form
factors in the $\gamma N N$-vertex, we chose the solution in which the
bare pseudovector form factor is given by Eq.~(23) of
Ref.\cite{Kon99}, a di-pole with half-width $\Lambda^2 = 1.28 \mbox {GeV}^2$.
We also did the calculation using the other choice of the bare $\pi N
N$-vertex, given by Eq.~(24) of Ref.\cite{Kon99}, with the half-width
$\Lambda^2 = 1.33 \mbox {GeV}^2$ (not shown). We found that the
results for the nucleon-photon form factors do not depend
significantly on the choice of the $\pi N N$-vertex.

Since the $\gamma N N$-vertex obeys the Ward-Takahashi identity, the
form factors $F_1^{\pm}(p^2)$ are uniquely determined, see
Eqs.~(\ref{eq:wtiFm},\ref{eq:wtiFp}), by the functions parametrizing
the nucleon propagator calculated in Ref.\cite{Kon99}. We checked
numerically that these are satisfied by the converged vertex. One of
the consequences of the Ward-Takahashi identity is that
$(F_1^{\pm})^s=(F_1^{\pm})^v$ and therefore $F_1^{\pm}=0$ for the
neutron-photon vertex. The form factors $F_1^{\pm}(p^2)$ in the
proton-photon vertex are depicted in \figref{f9}. They do not depend
on the choice of the $\gamma \pi N N$-vertex.

The dominant contribution to the form factors $F_2$ is due to
$\Gamma_\mu[1]$. Since this term is an inhomogeneity of the equation,
the bulk of the magnitude of the form factors is already generated in
the first iteration. This, however, does not mean that the other
integrals on the right-hand side of \eqref{sys} are of minor
importance. In particular, they are crucial for satisfying the
Ward-Takahashi identity.

In \figref{f2-p} the imaginary and real parts of the form factors
$F_2^+(p^2)$ (the solid line) and $F_2^-(p^2)$ (the dotted line) are
shown for the proton. The slope of $Im\,F_2^-(p^2)$ at the pion
threshold, $p^2=(m+m_{\pi})^2$, is much steeper as compared to that of
$Im\,F_2^+(p^2)$. As a consequence of this, we obtain a pronounced
cusp-like behavior of $Re\,F_2^-(p^2)$ at the threshold. The reason
is that in pion photoproduction the $E_{0^+}$ multipole has a finite
value at threshold while other multipoles tend to zero. Since this
multipole corresponds to spin and parity $1/2^-$ in the coupled
nucleon-photon channel it contributes to the imaginary part of $F_2^-$
which now obtains a term proportional to the 3-momentum of the cut
intermediate pion. The real part of $F_2^-$ calculated from a
dispersion integral thus exhibits a pronounced cusp structure,
contrary to the case of $F_2^+$ ($F_2^+$ is associated with the
positive energy component of the off-shell nucleon carrying
$J^\pi=1/2^+$). The magnitude of the cusp in $F_2^-$ depends thus on
the magnitude of the $E_{0^+}$ multipole multiplied by a weighted
difference of the pseudoscalar and pseudovector coupling strengths in the pion-nucleon
vertex.

The form factors in \figref{f2-p} are calculated using the $\gamma \pi N
N$ contact term of \eqref{gampinn1} when evaluating
$\Gamma_{\mu,I}[3]$ in \eqref{sys}.
An alternative form of the $\gamma \pi N N$-vertex is
obtained by using \eqref{contPV2} instead of \eqref{contPV1}. The
difference between these two contact terms, \eqref{contPVdiff}, is
transverse to the photon momentum. To illustrate the influence of the
different choices of the contact terms on the $\gamma N N$-vertex, in
\figref{f2-p}, right panel, we show the form factors $F_2^{\pm}(p^2)$ calculated using
the alternative contact term. From a comparison of left and right
panels in \figref{f2-p}, it
follows that the different choices of the contact term affect mainly
$F_2^-(p^2)$. The form factor $F_2^+(p^2)$ is
normalized at $p^2=m^2$ to the physical anomalous magnetic moment of
the nucleon and is only slightly sensitive to the choice of the
contact term. The difference between the two choices for the contact
terms shows most strongly in the $E_{0^+}$ multipole in
pion-photoproduction which explains why mainly $F_2^-(p^2)$ is
affected. In a full calculation one would fix the ambiguity in the
contact term from a
calculation of pion-photoproduction. This goes, however, beyond the
scope of the present work but will be discussed in a forthcoming
publication.

The results for the form factors $F_2^{\pm}(p^2)$ in the
neutron-photon vertex are shown in \figref{f2-n} for the two choices
of the $\gamma \pi N N$-vertex. The conclusions drawn above for the
proton apply qualitatively to this case as well.

The renormalization conditions, \eqref{normalF2}, are fulfilled by
adjusting the bare renormalization constants $\kappa_B^{s,v}$ defined
in \eqref{ver_bare}. We obtain $\kappa_B^s=0.03$ and $\kappa_B^v=1.51$
if the contact term \eqref{gampinn1} is used and $\kappa_B^s=0$ and
$\kappa_B^v=1.6$ for the alternative contact term.

The
off-shell form factors by themselves cannot unambiguously be extracted
from experiment. In
particular,
they can be changed by a redefinition of the nucleon field.
At the same time, the field redefinition will in
general also change the four- and higher-point vertices. It is known that the
S-matrix is independent of the representation of the fields \cite{kamef}.
Therefore, in a consistent calculation of the observables, two-
three- and higher-point Green's functions should be treated
using the same model assumptions and representation of the fields.
The link between off-shell effects
and contact interactions was emphasized in \cite{Sch95}, where
pion-nucleon scattering,
Compton scattering by a pion and bremsstrahlung processes were
considered. The form factors in the present approach are constructed
consistently with the nucleon self-energy and the K-matrix, using the
same representation to treat all these quantities. A certain care
should however be excersized in applying these form factors in other
calculations.

\section{Application in Compton scattering}

As an example of the application of the formalism, Compton scattering
off the nucleon will be calculated. Since only a very restricted model
space is included in the present calculation, we do not make a direct
comparison with experimental data. For a definitive comparison
with experiment, other important degrees of freedom, such as the
$\Delta$-resonance, would have to be included. This extension of the model
is in progress.

The amplitude for the Compton scattering process is obtained through
solving \eqref{k5} in a partial wave basis \cite{Kor98}.
The K-matrix matrix elements are constructed as a sum of tree-level
Feynman diagrams where, however, dressed vertices and propagators are
used. The tree-level diagrams for Compton scattering can be written as
the sum of three contributions depicted in \figref{dia-K1},
\beq
K_{\mu \nu}(q,k) =K_{\mu \nu}^s(q,k)+K_{\mu \nu}^u(q,k) +
K_{\mu \nu}^{c}(q,k) \;.
\eqlab{Kc}
\eeq
The incoming and outgoing nucleon momenta are $p$ and $p'$,
respectively. The momenta of the incoming and outgoing photons are
$k^\nu$ and $-q^\mu$ so that energy and momentum conservation reads
$p'=p+k+q$.
The pole contributions to Compton scattering are given by the s- and u-channel
diagrams. $K_{\mu \nu}^{c}(q,k)$ denotes the matrix element of the contact
term given by
\eqref{ct-me}. This term is added to obtain a gauge-invariant matrix
elements and is constructed using the minimal substitution procedure.
 Obeying gauge invariance is important for satisfying
low-energy theorems for the matrix elements.

The contact term \eqref{ct-me} is, however, not unique and a purely
transverse contribution may be added,
\beq
K_{\mu \nu}^{c'}(q,k)=\overline{u}(p')\,4i\,\hat{e}^2 \Bigg\{
F_{c}((p+k)^2) \, (e_4)_\mu (q,p'-q) \, (\overline{e_4})_\nu (k,p+k)  +
\Big[ \begin{array}{c}\mu \longleftrightarrow \nu \\q \longleftrightarrow k
\end{array} \Big] \Bigg\}\,u(p) \;,
\eeq
where the operator $(e_4)_\mu (q,p'-q)$ is given by \eqref{basis}
(with photon momentum $q$ and off-shell nucleon momentum $p'-q$). The
form factor is given by
\beq
F_{c}(p^2)= \frac{\mathcal{P}}{\pi}
\int_{(m+m_{\pi})^2}^{\infty}\! {dp^{\prime 2} \over p^{\prime 2}-p^2}
{ \Big[ \mbox{Im} \tilde{F}_{2}^{-} (p^{\prime 2}) \Big]^2 \over \mbox{Tr}
\Big[ \Lambda_{-} \, \tilde{\Sigma}_I(p^{\prime 2}) \Big] } \;,
\eeq
where $ \tilde{F}_{2}^{-} $ and $\tilde{\Sigma}_I$ are
calculated from expressions similar to those for the negative energy form
factor $F_2$ and the imaginary part of the nucleon self-energy, except that
in the pion vertex, \eqref{pinnver}, we have put $G_2=0$.
The reason for adding this term is to take into
account a contribution arising from cutting the pion line in the
``handbag'' loop diagram. The contribution from this diagram, generated in
the K-matrix procedure, gives rise to a sharply increasing imaginary
contribution to the Compton amplitude and thus to a pronounced cusp
structure in the corresponding real part.

We checked numerically that the calculated amplitudes obey current
conservation which was also proven analytically. In addition the
calculations obey constraints imposed by low-energy theorems.%\cite{gelgold,low}.

Since we work in a restricted model space we postpone a complete
comparison with experiment to a future publication where the model will
be extended to include the $\Delta$-isobar degree of freedom and
t-channel meson exchanges. Here we limit ourselves to the $f_{EE}^{1-}$
amplitude which shows a highly non-trivial behavior at the pion
production threshold.
Our results for the $f_{EE}^{1-}$ amplitude (the solid line in
\figref{EE1-}) shows a distinct cusp structure at the pion-production
threshold. This cusp is generated by the analyticity condition we
imposed. For comparison, the dashed line shows the partial wave
amplitude
calculated using the K-matrix built with the bare vertices and the
free nucleon propagator, $K_{c' c}=V_{c' c}$, thereby neglecting the
principal-value parts of the loop integrals contributing to the
T-matrix (see Eqs.~(\ref{eq:k3},\ref{eq:k4})) which does not show the
cusp. The effect of
unitarization on the real part of this amplitude are small,
indistinguishable in the figure. We also show the results extracted
from pion-photoproduction data through the application of analyticity
consideration\cite{Ber93}. A similar cusp structure is also seen in the
analysis of Ref.\cite{Hun97}. Extending our model with other degrees of
freedom will add a smooth function to $f_{EE}^{1-}$, changing the value
at the higher energies but should not affect the cusp structure.

Assuming that Compton scattering is dominated by dipole process, the
electric polarizability $\oln{\alpha}$ can be extracted from
$f_{EE}^{1-}$\cite{Ber93},
\beq
f_{EE}^{1-}=f_{B}+ { \oln{\alpha} \over 3} E_\gamma^2 \;,
\eeq
where $f_{B}$ is the Born amplitude. Using this relation we extract
$\oln{\alpha}=11.6\cdot 10^{-4} fm^3$, which is close to the value
obtained from chiral-perturbation theory\cite{Ber95}.

\Omit{
According to the low-energy theorem for Compton scattering \cite{gelgold,low},
the crossection for small photon energies $\omega$ can be written as
\beq
\frac{d\sigma}{d\Omega}_{lab}=\left(\frac{d\sigma}{d\Omega}\right)_{\!\!Born}
\!\!(1+  c_2 \omega^2+\ldots) \;,
\eqlab{let}
\eeq
interms of the Born cross section.
We have by an explicit analytical calculation verified that our
amplitude obeys this low-energy theorem.
Conventionally the  coefficient $c_2$ is expressed in terms of electric
($\oln{\alpha}$) and magnetic ($\oln{\beta}$) polarizabilities of the nucleon,
\beq
-\frac{m}{2\alpha}\Big[(\oln{\alpha}+\oln{\beta})(1+cos \theta)^2+
(\oln{\alpha}-\oln{\beta})(1-cos \theta)^2\Big].
\eqlab{c2_pol}
\eeq
with $\alpha \approx 1/137$, the electromagnetic coupling constant.
We have extracted these constants from
our full calculation presented in \figref{f12}.
The effect on the
polarizabilities due to the inclusion of
the principal-value parts of the nucleon-pion loop integrals in the
T-matrix is $\Delta\left(\oln{\alpha}+\oln{\beta}\right)=-2.4\cdot
10^{-4} fm^3$ and $\Delta\left(\oln{\alpha}-\oln{\beta}\right) =
3.3\cdot 10^{-4} fm^3$ showing clearly the dominant effect of
the form factors $F_2^{\pm}$.
}

\section{Conclusions}

We have presented a model for Compton
scattering on the nucleon. In this model special attention is
payed to observing analyticity in addition to unitarity, crossing
symmetry and gauge invariance. The model
is formulated in terms of half-off-shell form factors in the vertices
and a nucleon self-energy which carry the non-perturbative dressing due to the
non-pole contributions of pion-loop diagrams. The pole contributions are
taken into account through the use of a K-matrix formalism.

The key element of the model is an integral equation which describes
dressing of the $\gamma NN$-vertex with an infinite number of pion loops. In the
solution procedure we take advantage of unitarity and analyticity
considerations by using dispersion relations \cite{Bin60}
to obtain the real parts of the form factors from their imaginary parts.
The latter, in turn, are obtained by applying cutting rules \cite{cut},
with only the one-pion-nucleon discontinuities of the loop integrals
taken into account. The dependence of the form factors on the
four-momentum squared of the off-shell nucleon deviates from a monopole-
(or dipole-) like shape adopted often in phenomenological applications.
In particular, a characteristic feature of our results is a cusp-like
structure of the form factors in the vicinity of the one-pion threshold,
showing most clearly in the magnetic form factors corresponding to
negative-energy states of the off-shell nucleon.

One of the important requirements for the electromagnetic vertex is
obeying the Ward-Takahashi identity, which relates the vertex with the
nucleon propagator. We have included a four-point $\gamma
\pi N N$ term in our model to obey this condition. In a theory with
nucleon-pion form factors such a term is always necessary. We construct
a $\gamma \pi N N$-vertex using the prescription of minimal
substitution. Terms in the contact vertex which are transverse to the
photon momentum are not uniquely determined. As an example of this
ambiguity, we have constructed two contact terms with different
transverse components. We used these two contact terms in the
calculation of the half-off-shell nucleon-photon vertex and found that
the negative-energy magnetic form factors are influenced noticeably by
the choice of the contact term, while the effect on the positive-energy
form factors is rather small.

It should be emphasized that off-shell vertices (as any general Green's
functions, for that matter) depend not only on the model used to
calculate them, but also on the representation of fields in the
Lagrangian. In contrast, the measurable physical observables are
obtained from the scattering matrix and are therefore oblivious to the
representation of the Lagrangian (see, e.g., \cite{kamef,Sch95}). Even though
information on the half-off-shell vertices cannot be unambiguously
extracted from experiment, they are important for the calculation of
observables.

We have argued that the vertices and propagator generated in the
present dressing procedure are  consistent with a
coupled-channel K-matrix approach to Compton scattering, pion
photoproduction and pion scattering. We have shown effects of the dressing on the
cross section for real Compton scattering. An extension
of the model to include additional degrees of freedom is in progress.

\acknowledgements

This work is part of the research program of the ``Stichting voor
Fundamenteel Onderzoek der Materie'' (FOM) with financial support
from the ``Nederlandse Organisatie voor Wetenschappelijk
Onderzoek'' (NWO). We  would like to thank
Alex Korchin, Rob Timmermans and John Tjon for discussions.

\appendix

\section{Projection method}

The  calculation of the imaginary parts of the form
factors (the pole
contributions in \eqref{sys}) is formulated in terms of the following projection procedure.
The half-off-shell vertex \eqref{finon} can be
regarded as a vector in a four-dimensional linear space. For the sake of
generality we present here the procedure for a virtual photon where
the vertex is a vector in a six-dimensional space $V_6$. with the basis
\beq
\begin{array}{ll}
{\displaystyle (e_1)_{\mu} = \Lambda_{+}(p^{\prime}) \gamma_{\mu}
\Lambda_{+}(p)} \;, &
{\displaystyle (e_2)_{\mu} = \Lambda_{+}(p^{\prime}) \gamma_{\mu}
\Lambda_{-}(p) } \;, \\
{\displaystyle (e_3)_{\mu} = \Lambda_{+}(p^{\prime})
\, i\frac{\sigma_{\mu \nu} q^{\nu}}{2 m} \Lambda_{+}(p) } \;, &
{\displaystyle (e_4)_{\mu} = \Lambda_{+}(p^{\prime})
\, i\frac{\sigma_{\mu \nu} q^{\nu}}{2 m} \Lambda_{-}(p) } \;, \\
{\displaystyle (e_5)_{\mu} = \Lambda_{+}(p^{\prime}) \frac{q_{\mu}}{m}
\Lambda_{+}(p) } \;, &
{\displaystyle (e_6)_{\mu} = \Lambda_{+}(p^{\prime}) \frac{q_{\mu}}{m}
\Lambda_{-}(p) } \;,
\end{array}
\eqlab{basis}
\eeq
defined over a ring of complex-valued functions (form factors). For the
case of a real photon, as discussed in the present paper,
basis vectors $e_5$ and $e_6$ can be truncated from the space.
Thus, to find contributions to the imaginary parts of the form factors
$F_i^{\pm}$ from the integral $\Gamma_{\mu, I}[k], k=1\cdots 4$, amounts to finding the
coefficients of the expansion in the basis \eqref{basis}.

The dual space $V_6^{*}$ can be defined as spanned over the basis
$(\theta^i)^{\mu} = g^{\mu \lambda} \overline{(e_i})_{\lambda}$, where the
over-lining denotes the Dirac conjugate of an operator,
$\overline{A} \equiv \gamma_0 A^{\dagger} \gamma_0$.
\Omit{
Explicitly, we have
\beq
\begin{array}{ll}
{\displaystyle (\theta^1)^{\mu} = \Lambda_{+}(p) \gamma^{\mu}
\Lambda_{+}(p^{\prime}) } \;, &
{\displaystyle (\theta^2)^{\mu} = \Lambda_{-}(p) \gamma^{\mu}
\Lambda_{+}(p^{\prime}) } \;, \\
{\displaystyle (\theta^3)^{\mu} = \Lambda_{+}(p)
i \frac{\sigma^{\mu \nu} q_{\nu}}{2 m} \Lambda_{+}(p^{\prime}) } \;, &
{\displaystyle (\theta^4)^{\mu} = \Lambda_{-}(p)
i \frac{\sigma^{\mu \nu} q_{\nu}}{2 m} \Lambda_{+}(p^{\prime}) } \;, \\
{\displaystyle (\theta^5)^{\mu} = \Lambda_{+}(p) (-)\frac{q^{\mu}}{m}
\Lambda_{+}(p^{\prime}) } \;, &
{\displaystyle (\theta^6)^{\mu} = \Lambda_{-}(p) (-)\frac{q^{\mu}}{m}
\Lambda_{+}(p^{\prime}) } \;.
\end{array}
\eqlab{dbasis}
\eeq
}
We define the scalar product of
$\omega^{\mu} \in V_6^{*} $ and $v_{\mu} \in V_6$ as
\beq
\langle\, \omega^{\mu},\, v_{\mu}\, \rangle = \mbox{Tr} (\omega^{\mu} v_{\mu}) \;,
\eqlab{dual}
\eeq
with a tacit summation over $\mu$ and the trace taken in spinor space.
For the evaluation of the traces we used the algebraic programming
system  REDUCE \cite{red}.
 Now if
\beq
v_{\mu} = \sum_{i=1}^{6} c^i (e_i)_{\mu} \;,
\eqlab{expans}
\eeq
then the coefficients are obtained from the formula
\beq
c^k = \sum_{l=1}^{6}
(E^{-1})^k_l\, \Big\langle\, (\theta^l)^{\mu},\, v_{\mu}\, \Big\rangle \;,
\eqlab{coeff}
\eeq
where the matrix
$E^i_j = \langle\, (\theta^i)^{\mu},\, (e_j)_{\mu}\, \rangle$.
The coefficients $c^k$ are the form factors (or, more
precisely, contributions to the imaginary parts of the form factors).
Thus, we identify
\beq
\begin{array}{l}
c^1=Im F_1^{+},\;\;\;\;c^2=Im F_1^{-},\;\;\;\; c^3=Im F_2^{+} \;, \\
c^4=Im F_2^{-},\;\;\;\; c^5=Im F_3^{+},\;\;\;\;c^6=Im F_3^{-} \;.
\end{array}
\eqlab{identif}
\eeq

\section{Minimal substitution}

The minimal substitution in momentum space amounts to the following replacement
of the nucleon momentum,
$P_\mu \longrightarrow \widetilde{P}_\mu=P_\mu-\hat{e}A_\mu$, where
$P_\mu$ has to be considered as an operator
acting on the right and $\hat{e} = e \hat{e}_N \equiv e(1+\tau_3)/2$.
If in a given term $P_\mu$ is the rightmost operator and thus acts on
the field of the incoming nucleon, it gives $p_\mu$ which has
c-number components. Our procedure closely follows that of
Ref.\cite{Oht89}.
Throughout this appendix we assume that the electromagnetic field $A_\mu$
carries the four-momentum $q_\mu$ directed inwards the vertex,
$[P_\nu,A_\mu]=q_\nu A_\mu$.
We thus obtain
\beq
P^2 A_\mu=A_\mu (P+q)^2 =A_\mu (p+q)^2 =(p+q)^2 A_\mu \;,
\eqlab{p2_scpr}
\eeq
where for ease of writing the nucleon spinor fields have been omitted.
More generally, for any smooth function $f(p^2)$ one obtains
\beq
f(P^2) A_\mu=A_\mu\,f((p+q)^2).
\eqlab{f_scpr}
\eeq

Under the minimal
substitution, the nucleon momentum squared changes as
\beq
P^2 \longrightarrow \widetilde{P}^2=p^2-2\hat{e}A\cdot p-
\hat{e}A\cdot q +O(A^2)=p^2-\hat{e}\,A^\mu(2p_\mu+q_\mu)+O(A^2).
\eqlab{ms_p2}
\eeq
Collecting the coefficients of the terms linear in $A^\mu$ results in
the photon vertex. This procedure in indicated by the symbol
$\longmapsto$, i.e.\
\beq
p^2 \longmapsto -\hat{e}(2p_\mu+q_\mu) \;,
\eqlab{p2-mao}
\eeq
which reads that upon minimal substitution a $p^2$ term in an n-point
Green's function generates an (n-point+photon) Green's function
corresponding to the vertex $\Gamma_\mu= -\hat{e}(2p_\mu+q_\mu)$.

To generalize this for an arbitrary function
$f(p^2)$, we first consider the following combination:
\begin{eqnarray}
P^2\, \widetilde{P}^2&=&p^4-2\hat{e}q^2\,A\cdot p-2\hat{e}A\cdot p\, p^2-4\hat{e}
q\cdot p\, A\cdot p-\hat{e}q^2\, A\cdot q-\hat{e}A\cdot q\, p^2 \nonumber \\
&&-2\hat{e}A\cdot q \,q\cdot p+O(A^2)
=p^4-\hat{e}\,A^\mu(2p_\mu+q_\mu)(p+q)^2+O(A^2), \eqlab{ms_p4}
\end{eqnarray}
where \eqref{ms_p2} has been used. The next step is to find the result of the
minimal substitution in the monomials $p^{2n},\,n=1,2,3\ldots\;$. Using
Eqs.~(\ref{eq:p2_scpr},\ref{eq:ms_p2},\ref{eq:ms_p4}), we have
\begin{eqnarray}
P^{2n}\longrightarrow \widetilde{P}^{2n}&=&p^{2n}-[2\hat{e}A\cdot p+
\hat{e}A\cdot q]
\,\stackrel{n-1}{\overbrace{P^2\cdots P^2}}-
P^2\,[2\hat{e}A\cdot p+\hat{e}A\cdot q]\,
\stackrel{n-2}{\overbrace{P^2\cdots P^2}} \nonumber \\
&&-\ldots-\stackrel{n-1}{\overbrace{P^2\cdots P^2}}\,[2\hat{e}A\cdot p+
\hat{e}A\cdot q]+O(A^2) \nonumber \\
&=&p^{2n}-\hat{e}\,A^\mu(2p_\mu+q_\mu)\,{\displaystyle \sum_{m=0}^{n-1}}
(p+q)^{2m}\,p^{2(n-1-m)}+O(A^2)\;. \eqlab{ms_p2n}
\end{eqnarray}
The corresponding vertex is thus given by
\beq
p^{2n}\longmapsto -\hat{e}\,(2p_\mu+q_\mu)\,\frac{(p+q)^{2n}-p^{2n}}
{(p+q)^2-p^2} \;,
\eqlab{p2n-mao}
\eeq
where the identity
\beq
\sum_{l=0}^{k-1} x^{2l} y^{2(k-l-1)} = \frac{y^{2k}-x^{2k}}{y^2-x^2} \;,
\eqlab{telescope}
\eeq
has been used.
Since a generic function $f(p^2)$ can be formally expanded in powers of $p^2$,
\beq
f(p^2)=\sum_k a_k p^{2k} \;,
\eqlab{f_expans}
\eeq
we obtain
\beq
f(p^2)\longmapsto -\hat{e}\,(2p_\mu+q_\mu)\,\frac{f((p+q)^2)-f(p^2)}
{(p+q)^2-p^2} \;.
\eqlab{f-mao}
\eeq

Minimal substitution in $\vslash{p}$ results in
\beq
\vslash{p}\longmapsto -\hat{e}\gamma_\mu \;.
\eqlab{msPSl}
\eeq
Under minimal substitution the product $f(p^2)\vslash{p}$ changes as
\begin{eqnarray}
f(P^2)\pslash{P} &\longrightarrow&
f(\widetilde{P}^2)\widetilde{\pslash{P}}=f(\widetilde{P}^2)\vslash{p}
- f((p+q)^2) \, \hat{e}A^\mu \gamma_\mu + O(A^2)
\nonumber \\
&=&f(p^2)\vslash{p}-\hat{e}A^\mu(2p_\mu+q_\mu)\,
{\displaystyle \frac{f(p+q)^2)-f(p^2)}{(p+q)^2-p^2}}\,
\vslash{p}-\hat{e}A^\mu \gamma_\mu\,f((p+q)^2)+O(A^2) \;, %\eqlab{msFpsl}
\end{eqnarray}
and hence
\beq
f(p^2)\vslash{p}\longmapsto -\hat{e}\,(2p_\mu+q_\mu)\,\frac{f(p+q)^2)-f(p^2)}
{(p+q)^2-p^2}\,\vslash{p}-\hat{e}\,\gamma_\mu\,f((p+q)^2) \;.
\eqlab{fpsl-mao}
\eeq

Some other useful formulas are stated without proof,
\bea
g(p\cdot k) &\longmapsto& -\hat{e}\,k_\mu \frac{g((p+q)\cdot k)-g(p\cdot k)}
{q\cdot k} \;,
\eqlab{gpdq-mao} \\
\frac{1}{(p+k)^2-p^2} &\longmapsto& -\hat{e}\,\frac{2k_\mu}{[(p+k+q)^2-(p+q)^2]
[(p+k)^2-p^2]} \;,
\eqlab{den-mao} \\
f(p^2)\,g(p\cdot k)&\longmapsto& -\hat{e}\Big\{ (2p_\mu+q_\mu)
{\ds \frac{f((p+q)^2)-f(p^2)}{(p+q)^2-p^2}}\,g(p\cdot k) \nonumber \\
&& +k_\mu f((p+q)^2)\,
{\ds \frac{g((p+q)\cdot k)-g(p\cdot k)}{q\cdot k}} \Big\} \;,
\eqlab{fg-mao}\\
p_\mu &\longmapsto& -\hat{e}g_{\mu \nu}.
\eqlab{pmu-mao}
\eea
where $k$ is the momentum of an uncharged third particle and
$g(p \cdot k)$ is a generic function.

The formulas for minimal substitution in $P'$, the momentum
associated with the outgoing nucleon, are analogous to the
above, except that everywhere $q$ should be replaced by $-q$.

Please note that the terms generated by this minimal substitution
procedure are free from poles in the limit of $q\longrightarrow 0$.

\section{Vertices from minimal substitution}

\subsection{The $\gamma \pi N N$-vertex}

Minimal substitution in the $\pi NN$-vertex is discussed separately
for the pseudoscalar vertex,
\beq
\left(\Gamma_{\pi N N}^{ps}\right)_\alpha=\tau_\alpha \gamma^5 f(p^2) +
                f(p^{\prime 2}) \tau_\alpha \gamma^5  \;,
\eqlab{ps_oFFin}
\eeq
and the pseudovector vertex,
\beq
\left(\Gamma_{\pi N N}^{pv1}\right)_\alpha=\tau_\alpha \gamma^5 g(p^2) \vslash{p}
+ \vslash{p}^{\prime} g(p^{\prime 2}) \gamma^5 \tau_\alpha  \;.
\eqlab{pv_oFFin}
\eeq
The sum of these reduces for the half-off-shell vertex \eqref{pinnver}
for $f(p^2)=G_1(p^2) - G_2(p^2)-G_1(m^2)/2$ and $g(p^2)=G_2(p^2)/m$.
Minimal substitution in Eqs.~(\ref{eq:ps_oFFin},\ref{eq:pv_oFFin}) gives
\beq
\left( \Gamma_{\gamma \pi N N}^{ps}\right)_\alpha^\mu=
-\gamma^5 \Big\{ \tau_\alpha \hat{e}(2p^{\mu}+q^{\mu})
\frac{f((p+q)^2)-f(p^2)}{(p+q)^2-p^2}+\hat{e}\tau_\alpha
(2p^{\prime \mu}-q^{\mu})\frac{f((p^{\prime}-q)^2)-f(p^{\prime 2})}
{(p^{\prime}-q)^2-p^{\prime 2}} \Big\} \;,
\eqlab{contPS}
\eeq
and
\begin{eqnarray}
\left( \Gamma_{\gamma \pi N N}^{pv1}\right)_\alpha^\mu=
&&-\tau_\alpha \hat{e}\gamma^5\Big\{ (2p^{\mu}+q^{\mu})\vslash{p}
{\displaystyle \frac{g((p+q)^2)-g(p^2)}{(p+q)^2-p^2}}+
\gamma^\mu g((p+q)^2) \Big\} \nonumber \\
&&-\hat{e}\tau_\alpha \Big\{ (2p^{\prime \mu}-q^{\mu})
{\displaystyle \frac{g((p^{\prime}-q)^2)-g(p^{\prime 2})}
{(p^{\prime}-q)^2-p^{\prime 2}}}
\vslash{p}^{\prime}+\gamma^\mu g((p^{\prime}-q)^2)
\Big\}\gamma^5 \;, \eqlab{contPV1}
\end{eqnarray}
respectively where $p'=p+q$.

Minimal substitution in
\beq
\left( \Gamma_{\pi N N}^{pv2}\right)_\alpha=
\tau_\alpha \gamma^5 \vslash{p} g(p^2) +
g(p^{\prime 2}) \vslash{p}^{\prime} \gamma^5 \tau_\alpha  \;,
\eqlab{pv_off-2}
\eeq
gives a different contact term (because $\tilde{P}_\mu$ and $\tilde{P}^2$
do not commute),
\begin{eqnarray}
\left( \Gamma_{\gamma \pi N N}^{pv2}\right)_\alpha^\mu=
&&-\tau_\alpha \hat{e}\gamma^5\Big\{ (2p^{\mu}+q^{\mu})(\vslash{p}+\vslash{q})
{\displaystyle \frac{g((p+q)^2)-g(p^2)}{(p+q)^2-p^2}}+
\gamma^\mu g(p^2) \Big\} \nonumber \\
&&-\hat{e}\tau_\alpha \Big\{ (2p^{\prime \mu}-q^{\mu})
{\displaystyle \frac{g((p^{\prime}-q)^2)-g(p^{\prime 2})}
{(p^{\prime}-q)^2-p^{\prime 2}}}
(\vslash{p}^{\prime}-\vslash{q})+\gamma^\mu g(p^{\prime 2})
\Big\}\gamma^5 \;.
\eqlab{contPV2}
\end{eqnarray}
The difference between the vertices in \eqref{contPV1} and \eqref{contPV2} equals
\begin{eqnarray}
\Delta^\mu&=&-\tau_\alpha \hat{e}\gamma^5\Big\{ (2p^{\mu}+q^{\mu})\vslash{q}
{\ds \frac{g((p+q)^2)-g(p^2)}{(p+q)^2-p^2}}+
\gamma^\mu [g(p^2)-g((p+q)^2)] \Big\} \nonumber \\
&&-\hat{e}\tau_\alpha \Big\{ (2p^{\prime \mu}-q^{\mu})
{\ds \frac{g((p^{\prime}-q)^2)-g(p^{\prime 2})}
{(p^{\prime}-q)^2-p^{\prime 2}}}
(-\vslash{q})+\gamma^\mu [g(p^{\prime 2})-g((p^{\prime}-q)^2)] \Big\}\gamma^5,
\eqlab{contPVdiff}
\end{eqnarray}
which is orthogonal to the photon momentum, $q\cdot \Delta =0$.
This presents one example
of the known ambiguity in constructing such contact vertices:
terms orthogonal to the photon momentum are not uniquely determined
by the minimal substitution prescription.

\subsection{The $\gamma \gamma N N$-vertex}

As a first step, the $\gamma NN$-vertex needs to be constructed which
reduces to the appropriate half-off-shell vertex  and in addition obeys the
Ward identity. It is constructed
through minimal substitution in the inverse dressed nucleon propagator,
\beq
 S^{-1}(p)={1\over 2}[\alpha(p^2)\vslash{p}+\vslash{p}\alpha(p^2)]+\beta(p^2)
\eqlab{prop-inv-ms}
\eeq
where $\beta(p^2)= -\alpha(p^2)\xi(p^2)$.
We can use Eqs.~(\ref{eq:f-mao},\ref{eq:fpsl-mao}) to write the
nucleon-photon vertex obtained by the minimal substitution as
\beq
\Gamma^{min}_\mu(p^\prime,p) = \hat{e}_N{\Big\{} \frac{p^\prime_\mu+p_\mu}
{p^{\prime 2} - p^2}\,
[S^{-1}(p^\prime)-S^{-1}(p)] \nonumber \\
+ { \alpha(p^{\prime 2}) +  \alpha(p^2) \over 2}
[\gamma_\mu-\vslash{q}\,\frac{2p_\mu+q_\mu}{p^{\prime 2}-p^2}]
{\Big\}}.
%\eqlab{gam-min-wti}
\eeq
where $p'=p+q$. This vertex clearly satisfies the Ward-Takahashi
identity \eqref{wti_gen}.
In principle, both nucleons can be off-shell in this vertex.

A general form for the vertex can now be written as
\bea
\Gamma_\mu(p^{\prime},p) &=&
\Gamma^{min}_\mu(p^{\prime},p)+
[\vslash{q},\gamma_\mu]\{ F(p^2)(\vslash{p}-m)+H(p^2)\} \nonumber \\
&+& \{ (\vslash{p}^\prime -m) F(p^{\prime2}) +
H(p^{\prime 2})\}[\vslash{q},\gamma_\mu] \;.
\eqlab{verFg}
\eea
To obtain the half-off-shell vertex with the outgoing
on-shell nucleon, we apply \eqref{verFg} to a positive-energy spinor
$\oln{u}(p^\prime)$ on the left,
$\oln{u}(p^\prime)\vslash{p}^\prime=\oln{u}(p^\prime)m$. Equating the
resulting
half-off-shell vertex to \eqref{finon} the functions  $F(p^2)$ and
$H(p^2)$ can be determined,
\bea
(F)^{s,v}(p^2)=\frac{(F_2^{+})^{s,v}(p^2)-(F_2^{-})^{s,v}(p^2)
}{8m^2}
+ {\alpha(p^2)-\alpha(m^2) \over 8( p^2-m^2)}
\eqlab{fF2} \\
(H)^{s,v}(p^2)=\frac{2(F_2^{+})^{s,v}(p^2)-(F_2^{+})^{s,v}(m^2)}{8m}
+ {\alpha(p^2) m +\beta(p^2) \over 4(p^2-m^2)} + {\alpha(m^2)-1 \over
16m}
\;,
\eqlab{gF2}
\eea
where an analogue of the Gordon identity has been used in the form
\beq
\oln{u}(p^\prime)\,(p_\mu+p^\prime_\mu)=\oln{u}(p^\prime)\,
(\gamma_\mu \vslash{p}+
\gamma_\mu m-i\sigma_{\mu \lambda}q^\lambda) \;.
%\eqlab{p+ppr_1}
\eeq

To obtain the  contact $\gamma \gamma N N$-vertex we perform a minimal
substitution in \eqref{verFg}, with a second photon field carrying
a momentum $k$ and polarization index $\nu$.
Since both incoming $p$ and outgoing $p^\prime$ nucleons are on the mass shell in Compton
scattering, we need
only the matrix element of the contact $\gamma \gamma N N$-vertex between the
positive-energy spinors of the incoming and outgoing nucleons,
%\newpage
\begin{eqnarray}
K_{\mu,\nu}^{ct}(q,k)&=&
\oln{u}(p^\prime)\,\Gamma^{ct}_{\mu \nu}(q,k)\,u(p)=
\oln{u}(p^\prime)\,i\hat{e}^2\,
\Bigg\{ \frac{\alpha((p+q)^2)m+\beta((p+q)^2)}{2[(p+q)^2-m^2]} \nonumber \\
&& \times\Big[\frac{(p_\mu+p^\prime_\mu-k_\mu)(p_\nu+p^\prime_\nu+q_\nu)}
{(p+q)^2-m^2} - \frac{(p_\mu+p^\prime_\mu+k_\mu)(p_\nu+p^\prime_\nu-q_\nu)}
{(p+k)^2-m^2}+2g_{\mu \nu}\Big] \nonumber \\
&&+\frac{\alpha((p+q)^2)-\alpha(m^2)}{2[(p+q)^2-m^2]}\,
\Big[(p_\nu+p^\prime_\nu+q_\nu)\gamma_\mu +
(p_\mu+p^\prime_\mu-k_\mu)\gamma_\nu\Big]  \nonumber \\
&&+\frac{H((p+q)^2)-H(m^2)}{(p+q)^2-m^2}\,\Big[ [\vslash{k},\gamma_\nu]
(p_\mu+p^\prime_\mu-k_\mu)+
(p_\nu+p^\prime_\nu+q_\nu)[\vslash{q},\gamma_\mu] \Big] \nonumber \\
&&+F((p+q)^2)\Big[[\vslash{k},\gamma_\nu] \gamma_\mu+
\gamma_\nu [\vslash{q},\gamma_\mu]\Big] +
\Big[ \begin{array}{c}\mu \longleftrightarrow \nu \\q \longleftrightarrow k
\end{array} \Big]
 \Bigg\}\,u(p) \;,
\eqlab{ct-me}
\end{eqnarray}
where $p^\prime=p+k+q$ and the notation introduced in
Eqs.~(\ref{eq:prop-inv-ms},\ref{eq:fF2},\ref{eq:gF2}) has been used.
In \eqref{ct-me} $H=H^s + H^v$ (and analogously for F) since the
contact term vanishes for the neutron.
The contact term is explicitly crossing symmetric due to the last
term in \eqref{ct-me}.

\section{Gauge invariance of the method}

The Ward-Takahashi identity, a consequence of gauge invariance, imposes an
important constraint on the nucleon-photon vertex,
\beq
q\cdot \Gamma(p^\prime,p)=\hat{e}_N \big[ S^{-1}(p')-S^{-1}(p) \big],
\eqlab{wti_gen}
\eeq
with the photon momentum $q^\mu=p^{\prime \mu}-p^{\mu}$. In the
following we prove that the photon vertex obtained in our procedure
obeys the Ward-Takahashi identity \eqref{wti_gen}.

Initially we assume that the
$\gamma N N$-vertex on the right-hand
side of \eqref{sys} obeys the Ward identity. As a first step
we first construct a tree level pion-photoproduction amplitude
and prove its gauge invariance.
The amplitude is written as a sum of s- ,u-, and t-channel contributions and a contact
term (see \figref{dia-K1}),
\beq
K_\alpha^\mu = \sum_{i=s,u,t,c} K_{i,\alpha}^\mu.
\eqlab{pigam-prod}
\eeq
The incoming pion caries momentum $k$, the outgoing photon $-q$ while
$p$ and $p^\prime$ are the on-mass-shell momenta of the incoming and outgoing nucleons,
respectively ($p'=p+k+q$).
Contracting each term in \eqref{pigam-prod} with the photon momentum yields
\bea
q_\mu K^\mu_{s,\alpha} &=&
 \tau_\alpha \hat{e}_N \oln{u}(p^\prime)\Gamma^5(p^\prime-k)u(p) \;,
\eqlab{gauge_t1}\\
q_\mu  K^\mu_{u,\alpha} &=&
 \hat{e}_N \tau_\alpha \oln{u}(p^\prime) \overline{\Gamma^5}(p^\prime-q)u(p) \;,
\eqlab{gauge_t2}\\
q_\mu  K^\mu_{t,\alpha} &=&
 \tau_\beta (\hat{e}_\pi)_{\beta \alpha} \oln{u}(p^\prime)\gamma^5 g\,
u(p) \;,
\eqlab{gauge_t3}\\
q_\mu  K^\mu_{c,\alpha} &=& -\tau_\alpha \hat{e}_N
\oln{u}(p^\prime)\big[ \Gamma^5(p^\prime-k) - \gamma^5 g \big]u(p)\nonumber\\
&&- \hat{e}_N \tau_\alpha
\oln{u}(p^\prime)\big[ \overline{\Gamma^5} (p^\prime-q) + \gamma^5 g \big]u(p) \;,
\eqlab{gauge_t4}
\end{eqnarray}
where we have used \eqref{wti_gen} and the fact that for an on-shell nucleon
with momentum $p$,  $\oln{u}(p)S^{-1}(p)=0=S^{-1}(p)u(p)$.
We also used the normalization condition for the
nucleon-pion vertex with both nucleons on-shell,
$\oln{u}(p^\prime)\Gamma^5(m)u(p)=\oln{u}(p^\prime)\gamma^5 g\, u(p)$.
Adding Eqs.~(\ref{eq:gauge_t1}-\ref{eq:gauge_t4}) and using
$[\hat{e}_N,\tau_\alpha]= \tau_\beta (\hat{e}_\pi)_{\beta \alpha}$,
we obtain the desired result,
$
q_\mu K^\mu_{\alpha} =0$ .

The gauge invariance of this
pion-photoproduction amplitude is used to show that the solution of \eqref{sys} obeys
the Ward-Takahashi identity for the nucleon-photon vertex.
This can be done in a transparent way with the help of diagrammatic
expressions.
The pole contribution to the vertex is given by the sum of cut loop
diagrams entering in
the dispersion integral in \eqref{sys}.
(We assume here that the convergence of the
procedure has been reached.)
This sum can be rewritten by adding and
subtracting an additional diagram containing the pole contribution to the
self-energy in the incoming nucleon leg, as shown in \figref{f4} (top).
Index $I$ on the
left-hand side ({\it l.h.s.}) of this equation indicates that
only the pole contribution to the vertex is considered.
To evaluate the scalar product of the {\it r.h.s.} with the photon
momentum $q_\mu$, it is convenient to rewrite the equation as shown in
\figref{f4} (bottom).
Here, a common sub-diagram, which is a nucleon-pion vertex,
has been extracted from the {\it r.h.s.},
and the asterisk indicates that an integration is tacitly understood over
the phase space of the cut (on-shell) nucleon and pion lines.
Such separation of a sub-diagram is
consistent with the interpretation of Cutkosky rules as a unitarity condition
\cite{cut}. Note also that the Dirac spinor $\oln{u}(p^\prime)$ is
explicitly identified with the outgoing nucleon line. The sum of diagrams in the
brackets is the scattering amplitude
$K_\alpha^\mu$ for pion photoproduction considered above, which
is gauge invariant, i.e.\ $q_\mu K_\alpha^\mu=0$.
Therefore,
only the last diagram on the  {\it r.h.s.} contributes to
\beq
q_\mu \oln{u}(p^\prime) \Gamma_I^\mu(p)=
- q_\mu \,\oln{u}(p^\prime)\Gamma^\mu(p) S(p) \Sigma_I(p) =
\hat{e}_N \oln{u}(p^\prime) \Sigma_I(p) \;,
\eqlab{qdotrhs}
\eeq
where $\Sigma_I(p)$ stands for the pole contribution to the nucleon
self-energy and we have used \eqref{wti_gen}  for
the vertex on the {\it r.h.s.}.
\eqref{qdotrhs} corresponds precisely to the Ward identity for the
pole contribution of the vertex since the pole contribution to
$S_0^{-1}(p)=(\vslash{p}-m)$ is zero and
$S^{-1}(p)=S_0^{-1}(p)-\Sigma(p)$.

%%%%%%%%%%%%%%%%%%%%%%%%%%%%%%%%%%%%%%%%%%%%%%%%%%%%

\newpage
\begin{figure}
\epsfxsize 8 cm
\centerline{\epsffile[0 334 615 455]{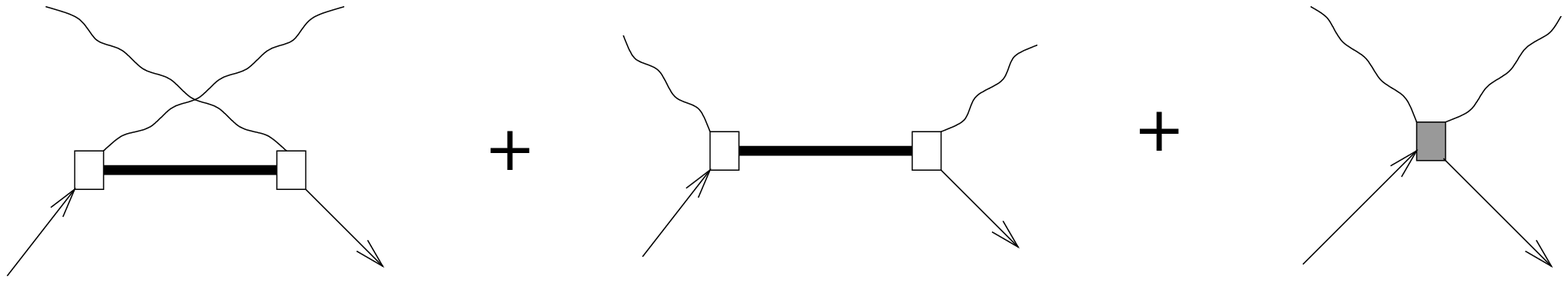}}
\epsfxsize 11 cm
\centerline{\epsffile[100 361 516 432]{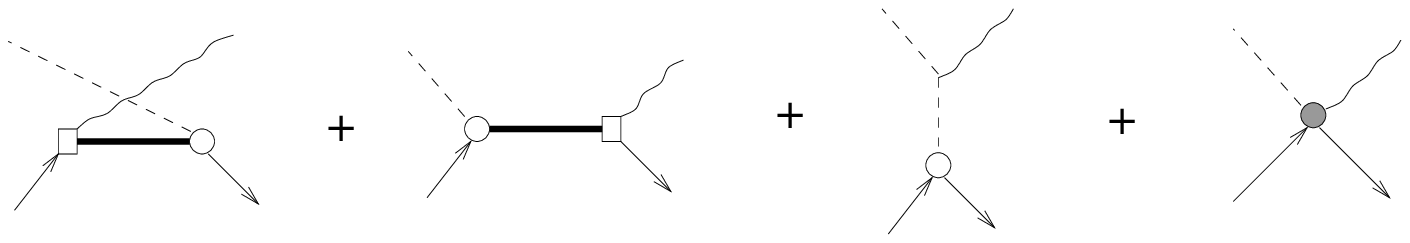}} %check
\caption[f10]{
The tree-level diagrams including the contact terms contributing to
the amplitude for Compton scattering (top) and pion
photoproduction (bottom). The solid, dashed and wavy lines
are nucleons, pions and photons, respectively. The thick line is the dressed
nucleon propagator. The empty circle (square) is the dressed $\pi N N$
($\gamma N N$)-vertex, and the shaded circle stands for the ``contact"
vertices. All the external lines are on-shell.
\figlab{dia-K1}}
\end{figure}

\begin{figure}
\epsfxsize 15 cm
\centerline{\epsffile[0 135 594 635]{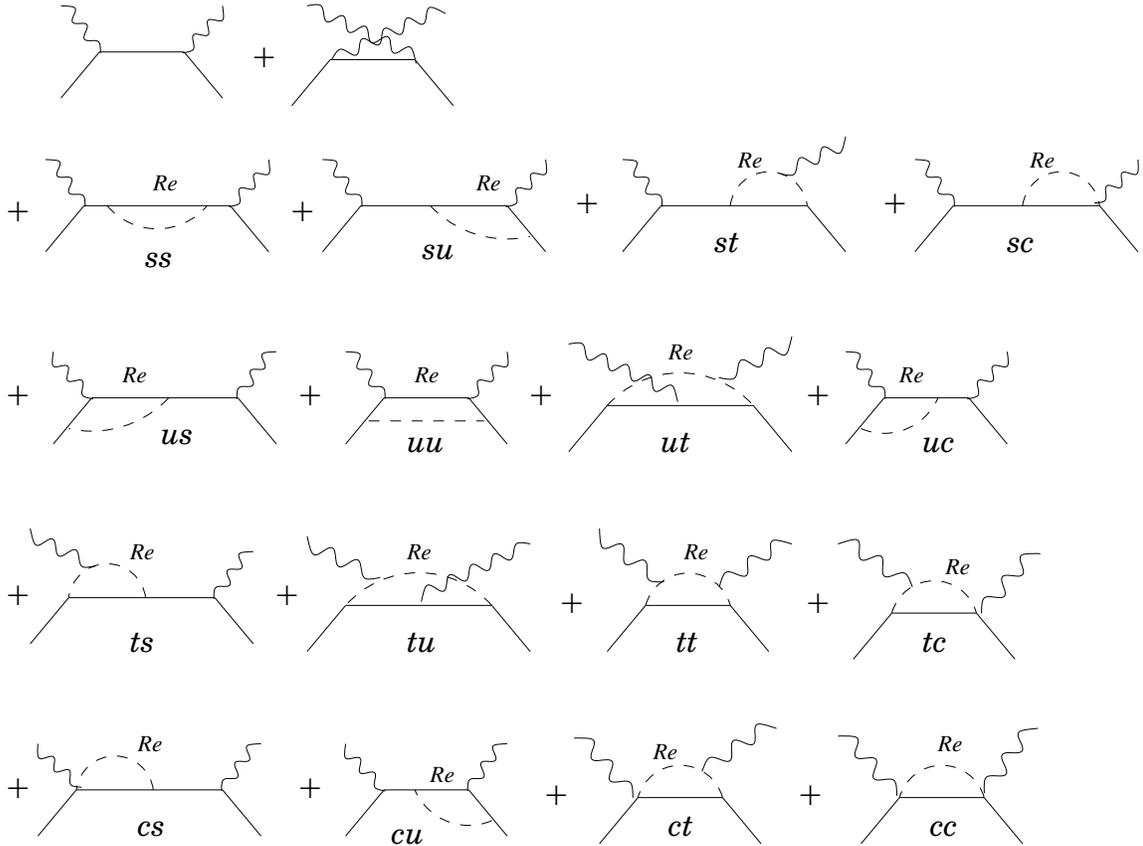}}
\caption[f11]{
The Feynman diagrams forming the K-matrix for Compton scattering up to second
order in the ``potential'' $V_{c' c}$, according to \eqref{k6}.
All the external lines are on-shell. The label {\it Re}
shows that
only the principal-value part of the loop integrals are included. The
structure of the loop diagrams is labeled in terms of the tree diagrams for pion
photoproduction, {\it ss, su, st} etc.
\figlab{f11}}
\end{figure}

\begin{figure}
\epsfxsize 15 cm
\centerline{\epsffile[-10 320 615 520]{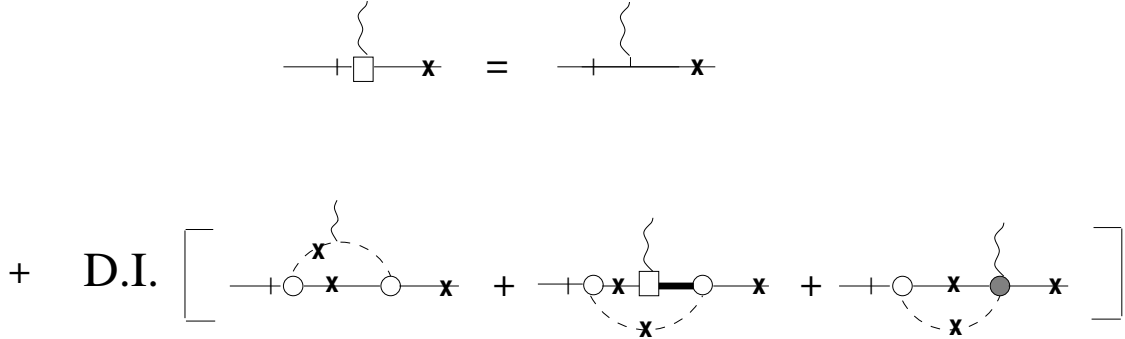}} %check
\caption[f1]{
The graphical representation of \eqref{sys} following the notation of
\figref{dia-K1}. Crosses denote cut propagators and cut external lines
are stripped away.
\figlab{f1}}
\end{figure}

\begin{figure}
\epsfxsize 8. cm
\centerline{\epsffile{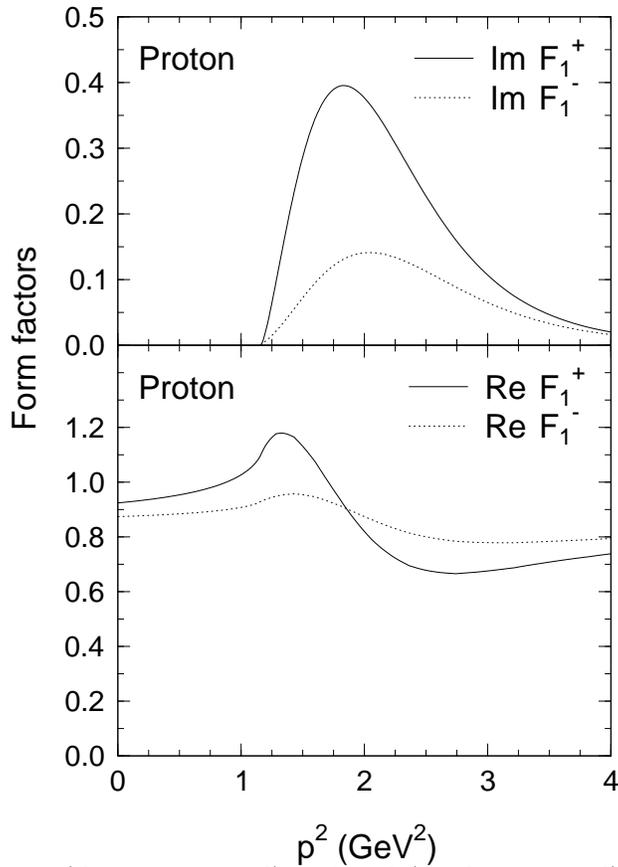}} %check
\caption[f9]{
The imaginary (the upper panel) and real (the lower panel) parts of the form
factors $F_1^{+}$ (the solid curves)
and $F_1^{-}$ (the dotted curves) as functions of the
momentum squared of the off-shell proton.
\figlab{f9}}
\end{figure}

\begin{figure}
\epsfxsize 10 cm
\centerline{\epsffile{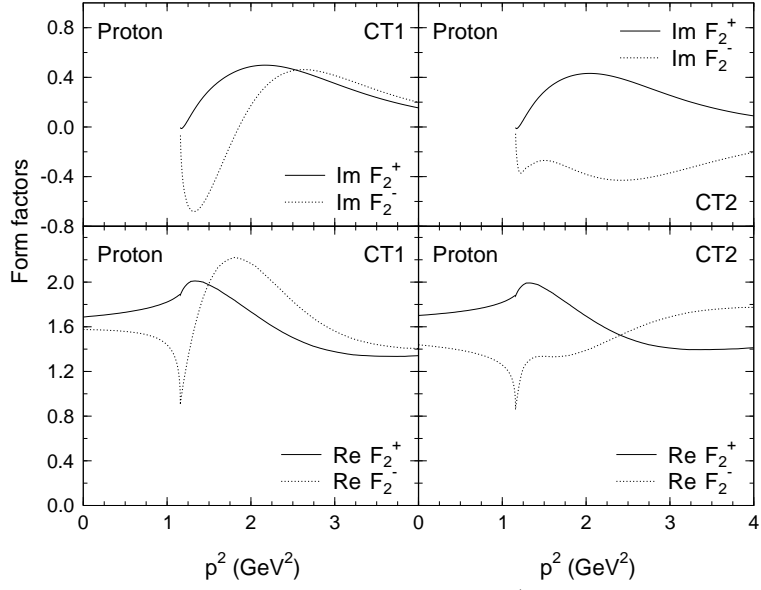}}
\caption[f5]{
The same as \figref{f9}, but for the form factors $F_2^{\pm}$ for two
different using $\gamma \pi N N$-contact terms, \eqref{gampinn1} (left panel) and
 \eqref{contPV2} (right panel).
\figlab{f2-p}}
\end{figure}

\begin{figure}
\epsfxsize 10. cm
\centerline{\epsffile{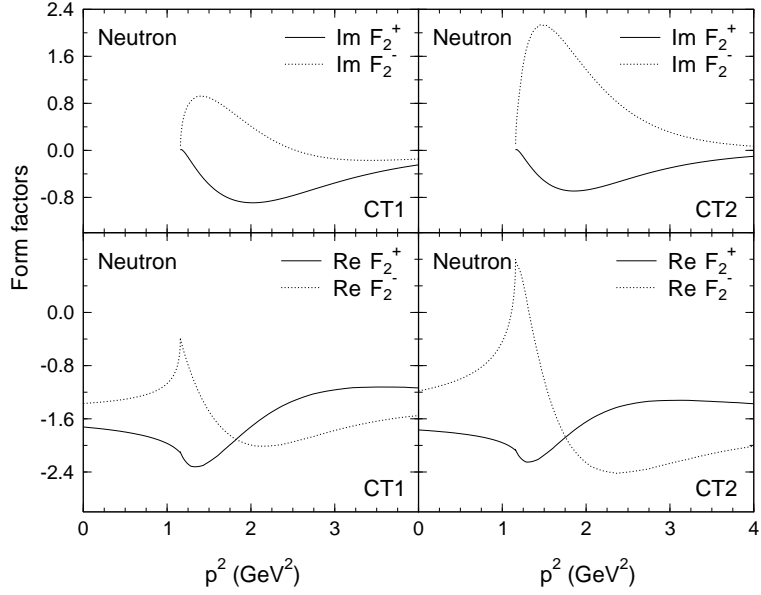}}
\caption[f7]{
The same as in \figref{f2-p}, but for the neutron-photon vertex.
\figlab{f2-n}}
\end{figure}

\begin{figure}
\epsfxsize 8 cm
\centerline{\epsffile[48 376 400 770]{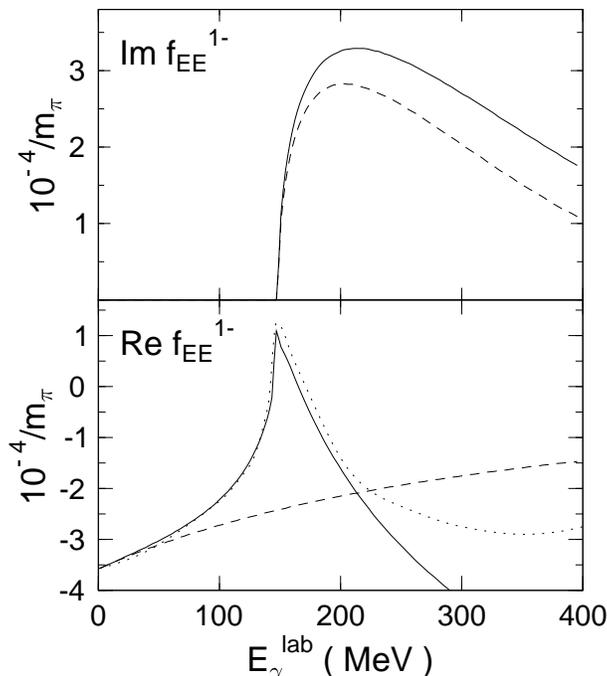}}
\caption[EE1-]{
The $f_{EE}^{1-}$ partial wave amplitude as function of the photon
energy.
The solid line represents the calculation where the dressed (irreducible)
nucleon-photon, nucleon-pion vertices and the dressed nucleon
propagator, are included in the K-matrix. The dashed line is the calculation
where the  bare vertices and the free nucleon
propagator are used. The dotted curve is the result based on an analysis
of pion-photoproduction data\cite{Ber93}.
\figlab{EE1-}}
\end{figure}

\begin{figure}
\epsfxsize 10 cm
\centerline{\epsffile[-10 150 630 670]{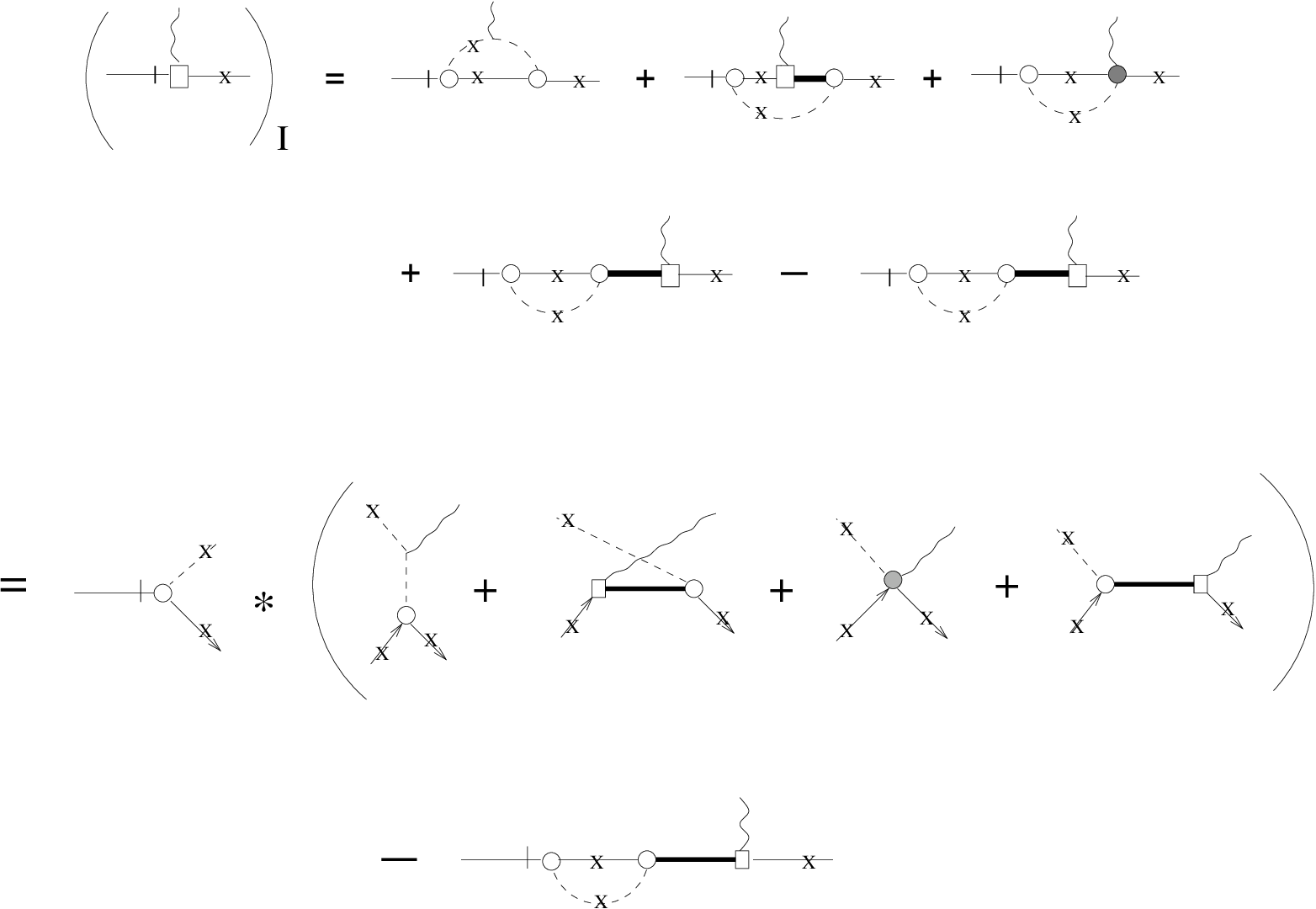}}
\caption[fig4]
{The diagrammatic equation for the pole contribution to the
$\gamma N N$-vertex (indicated by the subscript I) as used in the proof of the
consistency of the method with the Ward-Takahashi identity. The notation is as in
\figref{dia-K1}. The asterisk denotes an integration over the phase space of the
cut nucleon and pion lines.
\figlab{f4}}
\end{figure}

\end{document}